\documentclass[12pt]{iopart}
\usepackage{graphicx}

\begin{document}

\title{Supermassive black hole spin-flip during the inspiral}
\author{L\'{a}szl\'{o} \'{A}. Gergely$^{1,2\ast }$, Peter L. Biermann$%
^{3,4,5,6,7\dag }$, Lauren\c{t}iu I. Caramete$^{8\ddag }$ }

\address{
$^{1}$ Department of Theoretical Physics, University of Szeged, Hungary\\
$^{2}$ Department of Experimental Physics, University of Szeged, Hungary\\
$^{3}$ MPI for Radioastronomy, Bonn, Germany\\
$^{4}$ Department of Physics \& Astronomy, University of Bonn, Germany \\
$^{5}$ Department of Physics \& Astronomy, University of Alabama, Tuscaloosa, AL, USA\\
$^{6}$ Department of Physics, University of Alabama at Huntsville, AL, USA\\
$^{7}$ FZ Karlsruhe, and Physics Department, University of Karlsruhe, Germany\\
$^{8}$ Institute for Space Sciences, Bucharest, Romania \\ 
{ \small
$^{\ast }$ gergely@physx.u-szeged.hu;\  
$^{\dag }$ plbiermann@mpifr-bonn.mpg.de;\  
$^{\ddag }$ lcaramete@gmail.com
} }

\begin{abstract}
During post-Newtonian evolution of a compact binary, a mass ratio~$\nu $
different from 1 provides a second small parameter, which can lead to
unexpected results. We present a \textit{statistics of supermassive black
hole candidates}, which enables us first to derive their mass distribution,
then to establish a logarithmically even probability in $\nu $ of the mass
ratios at their encounter. In the mass ratio range $\nu \in \left(
1/30,1/3\right) $ of supermassive black hole mergers representing $40\%$ of
all possible cases, the combined effect of spin-orbit precession and
gravitational radiation leads to a spin-flip of the dominant spin during the 
\textit{inspiral} phase of the merger. This provides a mechanism for
explaining a large set of observations on X-shaped radio galaxies. In
another $40\%\,$ with mass ratios $\nu \in \left( 1/30,1/1000\right) $ a
spin-flip never occurs, while in the remaining $20\%$ of mergers with mass
ratios $\nu \in \left( 1/3,1\right) $ it may occur during the plunge. We
analyze the magnitude of the spin-flip angle occurring during the inspiral
as function of the mass ratio and original relative orientation of the spin
and orbital angular momentum. We also derive a formula for the \textit{final
spin }at the end of the inspiral in this mass ratio range.
\end{abstract}

\maketitle

\section{Introduction}

During galaxy mergers, following a regime of slow approach due to dynamical
friction, eventually the central supermassive black holes (SMBHs) approach
each other to a separation of the order of $10^{3}$ Schwarzschild radii,
when gravitational radiation takes over as the leading order dissipative
effect. The Laser Interferometer Space Antenna (LISA, see \cite{LISA}) is
expected to detect merging binary SMBHs with masses $m_{1}+m_{2}\le 10^{7}\,$%
solar masses (M$_{\odot }$) up to redshift $z\approx 30$. A post-Newtonian
approach is well suited to describe their forthcoming \textit{inspiral}, a
regime we define in terms of the post-Newtonian (PN) parameter $\varepsilon
=Gm/c^{2}r$ $\approx v^{2}/c^{2}\in \left( \varepsilon
_{in}=10^{-3},\varepsilon _{fin}=10^{-1}\right) $, where $r$ and $v$
characterize\ the orbital separation (from the center of mass) and speed of
the reduced mass particle, $G$ is the gravitational constant and $c$ the
speed of light. Various corrections to the conservative dynamics add up to 2
PN, while the gravitational radiation results in dissipation of energy,
angular momentum and orbital angular momentum at 2.5 PN.

The leading order conservative correction to the Newtonian dynamics in a
compact binary, which results in a change of the orbital plane (defined by
the direction $\mathbf{\hat{L}}_{N}$ of the Newtonian orbital angular
momentum $\mathbf{L}_{N}=\mu \mathbf{r}\times \mathbf{v}$ of the reduced
mass particle $\mu $) is the spin-orbit (SO) interaction \cite{SO}, \cite%
{ACST}. The precessional time-scale (the time during which the normal to the
orbit $\mathbf{\hat{L}}_{N}$ undergoes a full rotation) is longer than the
orbital period, however shorter than the characteristic time-scale of
gravitational radiation (defined as $L/\dot{L}$, where $L$ is the magnitude
of the total orbital angular momentum). Combined with the leading order
gravitational radiation backreaction averaged over one quasicircular orbit,
the SO correction provides a fair approximation to orbital dynamics,
explored in Refs. \cite{ACST}, \cite{SpinFlip}.

X-shaped radio galaxies (XRGs) exhibit two pairs of radio lobes and jets 
\cite{LeahyWilliams}, \cite{Xshape}. A recent review \cite{GBGW} summarizes
the four different models explaining XRGs: galaxy harbouring twin AGNs,
back-flow diversion models, rapid jet reorientation models, finally a new
jet-shell interaction model. A large subset of the observations (excepting
cases, when the jets are aligned with the optical axes of the host
ellipticals \cite{Battye}) are well-explained by the jet reorientation
model, which in turn implies a spin-flip \cite{LeahyWilliams}, \cite%
{Xspinflip} of the dominant black hole.

The details of how this would occur were worked out in Ref. \cite{SpinFlip}.
A key element was the determination of the typical mass ratio at SMBH
mergers by a series of estimates, which resulted in mass ratios $\nu
=m_{2}/m_{1}=1/30\,$\ to $\nu =1/3$. Because the spin scales with the mass
squared, the second spin was neglected and only the dominant spin $\mathbf{S}%
_{\mathbf{1}}$ (with magnitude $S_{1}$) kept. We summarize the consequences
of this model as follows.

a) For the typical mass ratio the dominance of $L$ over $S_{1}$ is reversed
as the separation in the binary decreases throughout the inspiral. In the
last stages of the inspiral the spin dominates over the orbital angular
momentum $S_{1}\gg L$.

b) The angle $\alpha $ between the orbital angular momentum and total
angular momentum $\mathbf{J}$ (with magnitude $J$), also the angle $\beta $
between the dominant spin and total angular momentum evolve as:%
\begin{eqnarray}
\dot{\alpha} &=&-\frac{\dot{L}}{J}\sin \alpha >0~,  \label{alphadot} \\
\dot{\beta} &=&\frac{\dot{L}}{J}\sin \alpha <0~.  \label{betadot}
\end{eqnarray}

c) The approximate expression relating $\alpha $ to the post-Newtonian
parameter $\varepsilon $, mass ratio $\nu $ and initial angle $\alpha +\beta 
$ span by the dominant spin with the orbital angular momentum (this angle
being a constant during the inspiral) is:%
\begin{equation}
\tan \alpha \approx \frac{\sin \left( \alpha +\beta \right) }{\varepsilon
^{-1/2}\nu +\cos \left( \alpha +\beta \right) }~.  \label{alpha}
\end{equation}%
(In Eq. (41) of Ref. \cite{SpinFlip} the left hand side was given as $\sin
2\alpha /\left( 1+\cos 2\alpha \right) $.)

In a criticism to the work presented in Ref. \cite{SpinFlip}, Gopakumar
recently argued that "it is unlikely that the spin-flip phenomenon will
occur during the binary black hole inspiral phase" \cite{GopakumarConf}.
This misconception comes from mixing up the \textit{instantaneous} change in
the direction of the total angular momentum, $\,d\mathbf{\hat{J}}/dt=\left( 
\dot{L}/J\right) \left[ \mathbf{\hat{L}}-\left( \mathbf{\hat{L}}\cdot 
\mathbf{\hat{J}}\right) \mathbf{\hat{J}}\right] \neq 0$ with its \textit{%
averaged} expression $\left\langle d\mathbf{\hat{J}}/dt\right\rangle =0$
over the precessional time-scale. The angles $\alpha $ and $\beta $ are 
\textit{not} constants during the post-Newtonian evolution, as claimed in
Ref. \cite{GopakumarConf}, they rather change as given in Eqs. (\ref%
{alphadot})-(\ref{betadot}).\footnote{%
Only when the total and orbital angular momenta are aligned, become the
angles $\alpha $ and $\beta $ individually constant, as they identically
vanish. Therefore in the aligned configuration no spin-flip could ever occur
by the combined mechanism of SO precession and gravitational radiation.}

In the present paper we revisit some of the arguments of the spin-flip
mechanism and also provide more details on it as compared to Ref \cite%
{SpinFlip}. In Section 2 we revisit the typical mass ratio argument,
following a recent statistics of supermassive black hole candidates,
resulting a newly established mass distribution. We comment on how these
findings would affect the typical mass ratio range at SMBH encounters. In
Section 3 we analyze how the spin-flip angle depends on the mass ratio and
relative orientation of the spin and orbital angular momentum. We also
derive a formula for the final spin during the inspiral. Finally we present
our Concluding Remarks.

\section{The sky in black holes: new statistics, consequences for the mass
ratio at SMBH encounters and chances of the spin-flip during the inspiral}

First we summarize the arguments of Ref. \cite{SpinFlip} on the mass ratios
at SMBH encounters. The mass distribution $\Phi _{BH}(M_{BH})$ of the
galactic central SMBHs in the mass range $3\times 10^{6}\div 3\times 10^{9}$
M$_{\odot }$ is well described by a power-law with an exponential cutoff,
but for our purposes can be adequately approximated by a broken power-law 
\cite{PressSchechter}-\cite{Lauer} (confirmed by an observational survey 
\cite{Ferrarese}). The break is at about $10^{8}$M$_{\odot }$. In agreement
with these arguments and observations we assume $\Phi _{BH}(M_{BH})\propto
M_{BH}^{-k}$, with $k\in \left( 1,2\right) $ below, and $\Phi
_{BH}(M_{BH})\propto M_{BH}^{-h}$, with $h\geq 3$\ above the break. Then the
probability for a specific mass ratio arose as an integral over the black
hole mass distribution, folded with the rate $\ F$ to merge, and by adopting
the lower values of the exponents. For the merger rate we assumed that it
scales with the capture cross section $S$ (the dependence on the relative
velocity of the two galaxies was neglected, as the universe is not old
enough for mass segregation). For the capture cross-section we assumed $%
S\propto \nu ^{-1/2}$, motivated by the following arguments:

\begin{itemize}
\item for galaxies an increase with a factor of $10$ in radius ($10^{2}$ in
cross-section) accounts for an increase with a factor of $10^{4}$ in mass
(from the comparison of our Galaxy with dwarf spheroidals \cite{Gilmore}-%
\cite{Klypin},

\item there is a well established correlation between the SMBH mass and the
mass of the host bulge \cite{Magorrian},

\item the mass of the central SMBH scales with both the spheroidal galaxy
mass component and the total, dark matter dominated mass of a galaxy \cite%
{Benson}.
\end{itemize}

As a result of these considerations we have found that most likely the mass
ratio is in the range $\nu \in \left( 1/30\,,1/3\right) $. A typical value
to consider would be $\nu =1/10$, thus one of the SMBHs being $10$ times as
massive as the other.

\begin{figure}[tbp]
\begin{center}
\includegraphics[width=15cm]{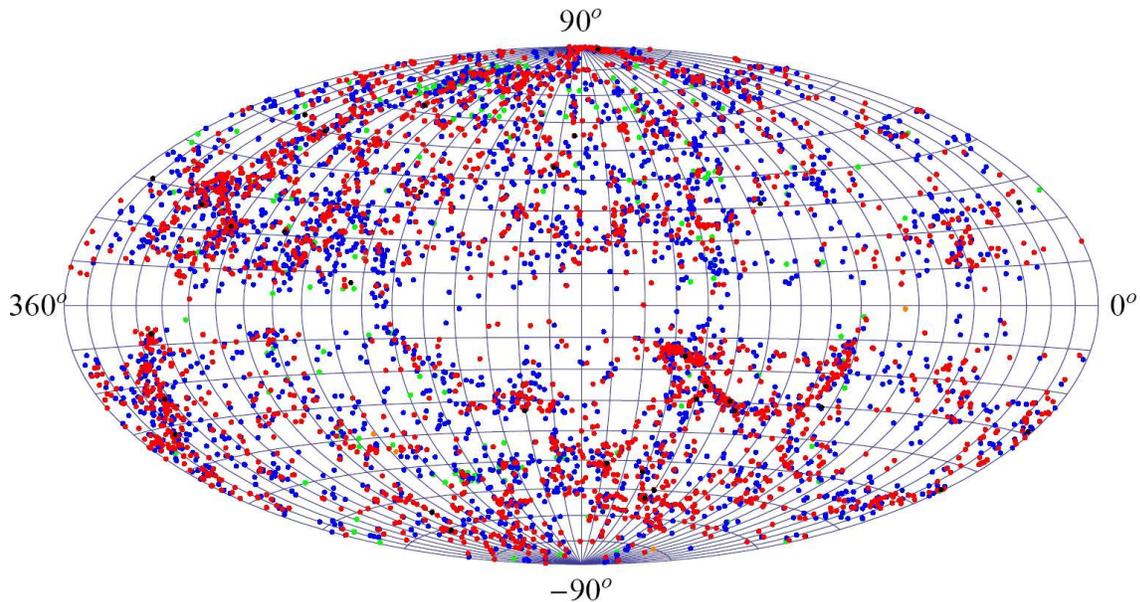}
\end{center}
\caption{Aitoff projection in galactic coordinates of 5,895 NED SMBH
candidate sources. The complete sample is complete in a sensitivity sense,
in order to derive densities one needs a volume correction. In the
electronic version the colour code is Orange, Green, Blue, Red, Black
corresponding to masses above $10^{5}$M$_{\odot }$, $10^{6}$M$_{\odot }$, $%
10^{7}$M$_{\odot }$, $10^{8}$M$_{\odot }$, $10^{9}$M$_{\odot }$,
respectively. With the exception of the less numerous first range (Orange),
representing compact star clusters, the rest are SMBHs.}
\label{Fig1}
\end{figure}

New work on the statistical analysis of 5,895 NED candidate sources \cite%
{CarameteBiermann} has been carried out in the mass range from $%
10^{5}\,M_{\odot }$ to above $10^{9}\,M_{\odot }$. Below about $%
10^{6}\,M_{\odot }$ all candidates are probably compact star clusters,
however the rest are likely SMBHs. This work shows that the SMBH mass
function is a broken power law with $M_{BH}^{-2}$ at low masses, and $%
M_{BH}^{-3}$ at high masses, with a break near $1.25\times 10^{8}\,M_{\odot }
$; this general behaviour has been long known, and has now been rederived
with a very large sample. The key difference with respect to previous work
was the careful attention paid in order to have equal probability for
detecting a SMBH in a galaxy, regardless to the Hubble type. The mass
distribution of the SMBHs is represented on Fig \ref{Fig1}. This particular
distribution can be interpreted in the context of the merger model \cite%
{SilkTakahashi} with a merger rate scaling as $(mass)^{+2}$, very much
stronger than what we favored in Ref \cite{SpinFlip}. The extreme mass
dependence describes well a $M_{BH}^{-3}$ black hole mass distribution
consistent with the high end of the mass distribution; on the other hand a
mass dependence of the merger rate close to $(mass)^{+4/3}$, suggested by
gravitational focusing arguments \cite{SilkTakahashi} describes well the
lower mass distribution nearer to $M_{BH}^{-2}$. It remains to be seen,
whether all details of the mass function can be understood using either of
these mass ratio dependences. Of course these simple merger rate
calculations assume an environment without cosmological expansion. However,
for the densest part of the cosmos the local expansion is very weak \cite%
{Cavaliere}, and that is where most of the mergers occur.

However, for the determination of the typical mass ratio the essential
result is only slightly changed. A merger rate running with mass$^{+2}$
analytically gives a $M_{BH}^{-3}$ mass function (see \cite{SilkTakahashi}),
as observed; we use this rate to estimate here the typical mass ratios for
high BH masses. Redoing the integrals of Section 2 of Ref. \cite{SpinFlip}
with $k=2$, $h=3$ (denoted there $\alpha ,\beta $) and $\xi =2$ (as in \cite%
{SilkTakahashi}, so much more extreme than what was assumed in \cite%
{SpinFlip}), then all four integrals are still dominated by the lower bound;
only the second of the integrals has $q\equiv \nu ^{-1}$ in its lower bound,
and so the four integrals have the $q$-dependencies of $q^{0}$, $q^{+1}$, $%
q^{-1}$ and again $q^{-1}$. We can ignore the second integral, since it all
refers to lower masses merging with lower masses. The most important
integrals are those combining a SMBH above the break with a SMBH either
below or above the break. Then the distribution in $q$ is found as $q^{-1}$,
a logarithmically even distribution in $(dq)/q$ over a range of $q$ from $1$
to $1000$, so a logarithmic average of $30$. Weighting the two parts of the
distribution, the larger mass ratios are favored, which would skew the
logarithmic average of the mass ratio to $q>\,30$, thus $\nu <0.03$.

The logarithmically even distribution means that the mass ratio ranges $\nu $
from $1$ to $\,1/3$, from $1/3$ to $1/10$, from $1/10$ to $1/30$, from $1/30$
to $1/100$, from $\,1/100$ to $1/300$ and finally from $\,1/300$ to $1/1000$
are roughly equal likely. A glance at Table 1 of Ref. \cite{SpinFlip} shows,
that concerning the behaviour of the ratio of the dominant spin and orbital
angular momentum magnitudes, we have three regimes:

(1) $\nu \in \left( 1/3,1\right) $ when $S_{1}<L\,\ $throughout the inspiral,

(2) $\nu \in \left( 1/30,1/3\right) $ when the initial $S_{1}<L\,\ $is
reversed to $S_{1}>L$ during the inspiral and

(3) $\nu $ $\in \left( 1/1000,1/30\right) $ when $S_{1}>L\,\ $holds
throughout the inspiral.

For the mass ratio ranges (1) and (3) no spin-flip can occur during the
inspiral, while for (2) it should. For (1) there is chance for a spin-flip
to occur during the plunge, as some numerical simulations have already found
this for equal masses \cite{Lousto}. For (3) by contrast there is no
possibility for a spin-flip by the combined mechanism of SO precession and
gravitational radiation. These mass ratio ranges then occur with (1) $20\%$,
(2) $40\%\,$\ and (3) again $40\%$ probability. This means that \textit{the
spin-flip still typically occurs during the inspiral}.

\section{Spin-flip angle distribution}

In this section we will present an analysis of the spin-flip angle occurring
during the inspiral phase in the mass ratio range $\nu \in \left(
1/30,1/3\right) $ as a function of the mass ratio.

The spin-flip model be understood as follows. Initially the galactic SMBH
has conserved spin, along which the primary jet can form. When the two
galaxies collide, the SO induced spin precession starts, while gravitational
radiation is diminishing the orbital angular momentum. The direction of the
total angular momentum stays unchanged. The constancy of $\mathbf{\hat{J}}$
over the precessional time-scale is due to the fact, that the change in the
total angular momentum $\mathbf{\dot{J}=}\dot{L}\mathbf{\hat{L}}$ is about
the orbital angular momentum, which (disregarding gravitational radiation)
undergoes a precessional motion about $\mathbf{J.}$ This shows that the
averaged change in $\mathbf{J}$ is along $\mathbf{J}$ (simple precession, 
\cite{ACST}). This conclusion, however, depends strongly on whether the
precessional angular frequency $\Omega _{p}$ is larger than $\dot{\alpha}$
and $\dot{\beta}$. Indeed, if these are comparable, the component
perpendicular to $\mathbf{J}$ in the change $\mathbf{\dot{J}=}\dot{L}\mathbf{%
\hat{L}}$ will not average out during one precessional cycle, as due to the
increase of $\alpha $ it can significantly differ at the beginning and at
the end of the same precessional cycle. Such a situation would occur, when
the spin and the orbital angular momentum are of comparable magnitude ($%
S_{1}\approx L$, a regime through which a binary with typical mass ratio
will pass through during the inspiral) and also roughly antialigned, a low
probability regime known as transitional precession. During simple
precession Eqs. (\ref{alphadot})-(\ref{betadot}) governing the evolution of
the angles $\alpha $ and $\beta $ also hold in an average sense over the
precessional time-scale. In what follows, we assume simple precession.

The magnitude of the spin is unaffected by gravitational radiation,
therefore by the simple rule of addition of vectors the spin has to align
close to the $\mathbf{\hat{J}}$ direction. The second jet then can start to
form. In the intermediate phase when the spin precesses, instead of jet
formation the precessing magnetic field creates a wind, sweeping away the
base of the old jet, which in many cases can be observed.

\subsection{Spin and orbital angular momentum orientations, final spin
formula}

The key equation to start with is Eq. (\ref{alpha}). In order to see the
validity of this equation, also to \textit{generalize }it to the cases of 
\textit{non-extreme rotation}, we need to evaluate

\begin{eqnarray}
S_{1} &\approx &m_{1}RV_{1}\approx m_{1}\frac{Gm_{1}}{c^{2}}c\frac{V_{1}}{c}%
\approx \frac{G}{c}m_{1}^{2}\chi _{1}~,  \nonumber \\
L &\approx &L_{N}\approx \mu rv=\frac{G}{c}\frac{v}{c}\frac{c^{2}r}{Gm}\mu m=%
\frac{G}{c}\varepsilon ^{-1/2}m_{1}m_{2}=\frac{G}{c}m_{1}^{2}\left(
\varepsilon ^{-1/2}\nu \right) ~.
\end{eqnarray}%
Here $V$ is some characteristic rotational velocity, $R$ the radius of the
SMBH (of the order of its Schwarzschild radius) and $\chi _{1}\in \left(
0,1\right) $ is the dimensionless ($\chi _{1}=1$\thinspace\ for extreme
rotation). Therefore 
\begin{equation}
\frac{S_{1}}{L}\approx \chi _{1}\varepsilon ^{1/2}\nu ^{-1}~.  \label{S1L}
\end{equation}%
Next we express $J/S_{1}$ first from $J=L\cos \alpha +S_{1}\cos \beta $ by
rewriting $\beta =\left( \alpha +\beta \right) -\alpha $, and secondly from
the equality of the projections perpendicular to $\mathbf{\hat{L}}$ of the
total and spin angular momenta $J\sin \alpha =S_{1}\sin \left( \alpha +\beta
\right) $, so that we can equal them. By also employing Eq. (\ref{S1L}) and
basic trigonometry we obtain

\begin{equation}
\tan \alpha \approx \frac{\sin \left( \alpha +\beta \right) }{\chi
_{1}^{-1}\varepsilon ^{-1/2}\nu +\cos \left( \alpha +\beta \right) }~.
\label{alphachi1}
\end{equation}
This is the desired generalization of Eq. (\ref{alpha}).

It is worth to note that when applied to the final configuration $%
\varepsilon _{fin}$, Eq. (\ref{alphachi1}) also stands as a formula for the 
\textit{final spin at the end of the inspiral}, giving the polar angle of
the final spin $\alpha _{fin}$ with respect to the axis $\mathbf{\hat{J}}$
in terms of the mass ratio, spin magnitude and angle span with the orbital
angular momentum. Related formulae based on numerical runs were advanced in
Refs. \cite{finalspin}. These results are not immediate to compare with
ours, as Eq. (\ref{alphachi1}) could at most be applied at the end of the
inspiral; although in the mass ratio range where it is valid, one would
intuitively expect that as not much orbital angular momentum is left at the
end of the inspiral in comparison with the dominant spin, the direction of
the latter will not be significantly changed during the plunge.

\subsubsection{Particular cases.}

There are three particular configurations worth to mention:

i) The spin is aligned with the orbital angular momentum: $\alpha +\beta =0$%
, thus from Eq. (\ref{alphachi1}) $\alpha =0$ and there is no room for any
spin-flip. This would be the situation for perfectly wet mergers, which
align the spin with the orbital angular momentum.

ii) The spin is anti-aligned with the orbital angular momentum, $\alpha
+\beta =\pi $. Therefore depending on which of the $S_{1}$ and $L$ are
larger, the angle $\alpha $ is either $0$ or $\pi $.

iii) For the parameter ranges when the denominator vanishes $\alpha +\beta
=\arccos \left( -\chi _{1}^{-1}\varepsilon ^{-1/2}\nu \right) $, from Eq. (%
\ref{alphachi1}) we obtain $\alpha =\pi /2$, therefore $\beta $ is also
determined.

\subsubsection{Discussion as function of mass ratios.}

Keeping in mind that due to Eqs. (\ref{alphadot})-(\ref{betadot}) the angle $%
\alpha +\beta $ is a constant during the inspiral (a parameter), and the
dimensionless spin $\chi _{1}$ behaves similarly, also regarding the mass
ratio as a third parameter characterizing the particular merger, the angle $%
\alpha $ in general remains a function of $\varepsilon $, thus it evolves
together with the orbital separation $r$ and velocity $v$.

For the mass ratio $\nu =1/10$ we have $\left( S_{1}/L\right) _{in}\approx
\chi _{1}\varepsilon _{in}^{1/2}\nu ^{-1}=0.316\chi _{1}$ and $\left(
S_{1}/L\right) _{fin}\approx \chi _{1}\varepsilon _{fin}^{1/2}\nu
^{-1}=3.162\chi _{1}$. As $\tan \alpha \leq S_{1}/L$ and $\tan \beta \leq
L/S_{1}$ (the equalities arising when the spin and orbital angular momentum
are perpendicular) we have $\tan \alpha _{in}\leq 0.316\chi _{1}$ and $\tan
\beta _{fin}\leq 0.316\chi _{1}^{-1}$. For extreme rotation ($\chi _{1}=1$)
we obtain $\alpha _{in},\beta _{fin}\leq 0.316=18.105^{\circ }$.

For $\nu =1/3$ we obtain $\tan \alpha _{in}\leq \left( S_{1}/L\right)
_{in}\approx \chi _{1}\varepsilon _{in}^{1/2}\nu ^{-1}=0.095\chi _{1}$ and
for extreme rotation $\alpha _{in}\leq 0.095=5.44^{\circ }$. In fact at the
beginning of the inspiral this latter condition (meaning that the orbital
angular momentum is roughly the total angular momentum) holds in the whole
mass ratio range $\nu \in \left( 1/3,1\right) $. Under these conditions Eq. (%
\ref{alphachi1}) can be approximated as%
\begin{equation}
\alpha _{in}\approx \chi _{1}\varepsilon _{in}^{1/2}\nu ^{-1}\sin \beta
_{in}=0.032\chi _{1}\nu ^{-1}\sin \beta _{in}~.
\end{equation}

For $\nu =1/30$ we obtain $\tan \beta _{fin}\leq \left( L/S_{1}\right)
_{fin}\approx \chi _{1}^{-1}\varepsilon _{fin}^{-1/2}\nu =0.105\chi _{1}^{-1}
$ and for extreme rotation $\beta _{fin}\leq 0.105=6.04^{\circ }$. In fact
at the beginning of the inspiral this latter condition (meaning that the
dominant spin is roughly the total angular momentum) holds in the whole
range $\nu \in \left( 1/1000,1/30\right) $. Under these conditions Eq. (\ref%
{alphachi1}) can be expanded (to first order in $\beta _{fin}$, with $\chi
_{1}^{-1}\varepsilon _{fin}^{-1/2}\nu $ of the order of $\beta $) as%
\begin{equation}
\beta _{fin}\approx \chi _{1}^{-1}\varepsilon _{fin}^{-1/2}\nu \sin \alpha
_{fin}=3.162\chi _{1}^{-1}\nu \sin \alpha _{fin}~.
\end{equation}%
For slowly rotating SMBHs with $\chi _{1}\approx 0.1$ the above formula
would hold only in the range $\nu \in \left( 1/1000,1/300\right) .$

\subsection{The spin-flip angle during the inspiral}

A minimal value for the spin-flip angle $\sigma $ arises by forming the
difference between the angles $\beta $, characterizing the orientation of
the spin with respect to the inertial direction $\mathbf{\hat{J}}$. Thus 
\begin{equation}
\sigma _{\min }=\beta _{in}-\beta _{fin}=\alpha _{fin}-\alpha _{in}~.
\label{sigmamin}
\end{equation}%
In the second equality we have used that $\alpha _{in}+\beta _{in}=\alpha
_{fin}+\beta _{fin}$.

However we have to take into account, that the above is only true in a
2-dimensional picture. In reality the 3-dimensional SO precession will
complicate the situation, and the above angle emerges only if the number of
precessions during the inspiral is an integer multiple of $2\pi $. If
instead is of the type $\left( 2k+1\right) \pi $ the spin-flip angle will be
maximal, to be calculated as%
\begin{equation}
\sigma _{\max }=\beta _{in}+\beta _{fin}-l\pi =2\left( \alpha _{in}+\beta
_{in}\right) -l\pi -\left( \alpha _{in}+\alpha _{fin}\right) ~,
\label{sigmamax}
\end{equation}%
where $l=0$ if $\beta _{in}+\beta _{fin}\leq \pi $ and $l=1$ if $\pi <\beta
_{in}+\beta _{fin}<2\pi $.

The difference between $\sigma _{\max }$ and $\sigma _{\min }$ is due to the
fact, that the realignment of the spin along $\mathbf{\hat{J}}$ is not
perfect. The closer $\mathbf{S}_{\mathbf{1}}^{fin}$ is to $\mathbf{\hat{J}}$%
, the less their difference ought to be due a more perfect alignement,
therefore $\sigma _{\max }-\sigma _{\min }=\beta _{fin}-\beta _{in}$ should
go to $0$ with decreasing $\nu .$

\begin{figure}[tbph]
\begin{center}
\includegraphics[width=12cm]{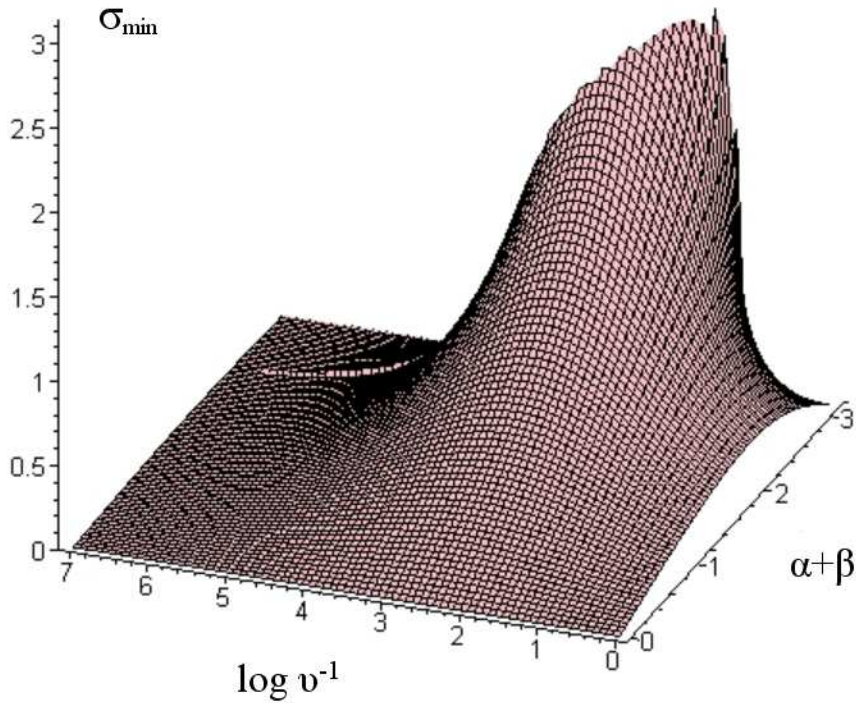}
\end{center}
\caption{The spin-flip angle $\protect\sigma _{\min }$ as function of the
relative orientation of the spin and orbital angular momentum $\protect%
\alpha +\protect\beta $ (a constant during inspiral), and mass ratio $%
\protect\nu $. For a given mass ratio the spin-flip angle has a maximum
shifted from $\protect\pi /2$ towards the anti-aligned configurations. The
mass ratios $\protect\nu =1;~1/3;~1/30$ and $1/1000$ are located on the $%
\log \protect\nu ^{-1}$ axis at $0;~1.09;~3.40$ and $6.91$, respectively,
confirming the prediction, that a significant spin-flip will occur in the
mass ratio range $\protect\nu \in \left( 1/30,1/3\right) $. For mass ratios
smaller than $1/100$ the spin does not flip at all, as the infalling SMBH
acts as a test particle.}
\label{Fig2}
\end{figure}

For generic mass ratios $\nu \in \left( 1/30,1/3\right) $, Eqs. (\ref%
{alphachi1}), (\ref{sigmamin}) and (\ref{sigmamax}) give the range of
allowed spin-flip angle for each relative orientation $\alpha +\beta $ of
the spin with respect to the plane of motion and each $\chi _{1}$. The
generic numerical solution for $\sigma _{\min }$ in the case $\chi _{1}=1$\
is represented on Fig \ref{Fig2} as function of the relative orientation of
the spin and orbital angular momentum $\alpha +\beta $ and mass ratio $\nu $%
. For a given mass ratio the spin-flip angle has a maximum shifted from $\pi
/2$ towards the anti-aligned configurations. The figure confirms the
prediction, that significant spin-flip will occur during the inspiral in the
mass ratio range $\nu \in \left( 1/30,1/3\right) $. For mass ratios smaller
than $1/100$ the spin does not flip at all, as the infalling SMBH acts as a
test particle. 

\section{Concluding Remarks}

In light of the new data on a large sample of SMBH candidates we have
established that the mass ratios obey an even logarithmic distribution in $%
\nu $. In the mass ratio range $\nu \in \left( 1/30,1/3\right) $ of SMBH
mergers representing $40\%$ of all possible cases, we have investigated the
SO precession driven conservative and gravitational radiation driven
dissipative contributions to the orbital evolution during the inspiral,
averaged over the precession time-scale. In this mass range the ratio of the
dominant spin magnitude and orbital angular momentum magnitude $S/L$ changes
from \textit{less than 1} to \textit{larger than 1} during the inspiral. As
the direction of the total angular momentum is unchanged on all time-scales
larger than the precession time-scale, while the magnitude of the the
orbital angular momentum decreases due to gravitational radiation and the
magnitude of the spin stays constant, the spin direction has to change. The
spin-flip of the dominant spin therefore occurs during the \textit{inspiral}%
. If jet activity is involved, X-shaped radio galaxies arise by this
mechanism and a large set of observations on X-shaped radio galaxies could
be explained.

In another $40\%\,$ of the mergers with mass ratios $\nu \in \left(
1/1000,1/30\right) $ the spin-flip never occurs by this mechanism, while in
the remaining $20\%$ of mergers with mass ratios $\nu \in \left(
1/3,1\right) $ it may occur during the plunge.

SMBH mergers of equal mass to $\nu=1/3$ are only half as likely as the mass
ratios $1/30$ to $1/3$, therefore the occurrence of the spin-flip can be
considered typical during the inspiral. We analyzed the magnitude of the
spin-flip angle occurring during the inspiral as function of the mass ratio
and original relative orientation of the spin and orbital angular momentum
and supported by numerical analysis the theoretical prediction (Fig \ref%
{Fig2}). We also derived a formula for the \textit{final spin }at the end of
the inspiral in this mass ratio range.

During the inspiral the following relations among the relevant time-scales
hold: tilt / spin-flip time-scale $\ge $ inspiral time-scale $\gg $
precession time-scale $\gg $ orbital time-scale (for all mass ratios in the
typical range). Interestingly enough, the spin-flip time-scale for a typical
mass ratio of $1/10$ is only about three years, while the precession
time-scale is less then a day \cite{SpinFlip}. Thus rapidly rotating
relativistic jets coming close to our line of sight could produce
significant variability at all wavelengths years before the coalescence.
Therefore electromagnetic counterparts / precursors to the strongest
gravitational wave emission are also likely to occur.

\section*{Acknowledgements}

This work was supported by the Pol\'{a}nyi Program of the Hungarian National
Office for Research and Technology (NKTH), the Hungarian Scientific Research
Fund (OTKA) grant 69036 and the COST Action MP0905 (L\'{A}G); AUGER
membership and theory grant 05 CU 5PD 1/2 via DESY/BMBF and VIHKOS (PLB);
and contracts CNCSIS 539/2009 and CNMP 82077/2008 (LIC).

\end{document}